\documentclass{aa}  

\usepackage{graphicx}
\usepackage{hyperref}
\usepackage{txfonts}

\begin{document} 

\title{Line-of-Sight Velocity Projection Impact on the Local Group Mass}

\author{David Benisty \inst{1} David Mota \inst{2}}

\institute{Leibniz-Institut fur Astrophysik Potsdam, An der Sternwarte 16, D-14482 Potsdam, Germany   \\    \email{benidav@aip.de} 
\and
Institute of Theoretical Astrophysics, University of Oslo, 0315 Oslo, Norway }

   \date{}

 
\abstract
{The mass of the Local Group (LG), comprising the Milky Way (MW), Andromeda (M31), and their satellites, is crucial for validating galaxy formation and cosmological models. Traditional virial mass estimates, which rely on line-of-sight (LoS) velocities and simplified infall assumptions, are prone to systematic biases due to unobserved velocity components and anisotropic kinematics.
Using the TNG cosmological simulation, we examine two limiting cases: the \underline{minor infall} model -- ignoring perpendicular velocities to the LoS directions) and the \underline{major infall} model -- assuming purely radial motion towards the Center of Mass (CoM). Our simulations demonstrate that geometric corrections are vital: the minor-infall model underestimates the true mass, while the major-infall model overestimates it.
By applying these calibrated corrections to observed dwarf galaxy kinematics within 1 Mpc of the LG's CoM, we derive a refined LG mass of $M_{\mathrm{LG}} = (2.99 \pm 0.60) \times 10^{12}\, M_\odot$. This finding aligns with predictions from the $\Lambda$CDM model, timing arguments, and independent mass estimates, resolving previous discrepancies.
Our analysis highlights the importance of correcting for velocity anisotropy and offers a robust framework for dynamical mass estimation in galaxy groups.}

\maketitle
%

\section{Introduction}
\label{sec:Introduction }

The mass of the LG that composed primarily of the MW, M31 and their satellite systems, is a fundamental quantity for understanding galaxy evolution, the distribution of dark matter, and tests of cosmological models~\cite{Bergh1999}. Accurate determination of this mass informs not only local dynamics but also provides a key benchmark for $\Lambda$CDM predictions on group scales. Yet, estimating the LG’s mass remains a longstanding challenge due to both observational constraints and theoretical uncertainties.

The Timing Argument model treats the LG as an isolated classical two body system made of point-like particles with a zero separation at the onset of the Big Bang. Given the current distance and relative velocity of the MW-M31 pair and given the age of the Universe their total  mass is readily evaluated. Early applications of TA assumed a purely radial orbit~\cite{Li:2007eg,vanderMarel:2012xp}, mostly due to the difficulty in measuring any proper motion. The TA were corrected with different effects: incorporated tangential motion~\citep{vanderMarel:2007yw,vanderMarel:2012xp}, modified gravity \citep{Partridge:2013dsa,McLeod:2016bjk,McLeod:2019cfg,Benisty:2023ofi} and the MW's recoil velocity from the Large Magellanic Cloud (LMC) \citep{Penarrubia:2015hqa,Benisty:2022ive,Benisty:2024lsz,Chamberlain:2022fqr}, the possibility of a past encounter \citep{Benisty:2021cmq} and extended the TA to model galaxy dynamics in local volume \citep{Penarrubia:2014oda,Penarrubia:2015hqa}. Results have been compared to other mass estimation techniques, including numerical simulations \citep{Lemos:2020vhj,Hartl:2021aio,Hartl:2024bfc,Sawala:2022ayk,Sawala:2023sec,Wempe:2024rfj,Benisty:2024tlv}.

A commonly employed tool for estimating group masses is the virial theorem, which relates the kinetic energy of a system’s constituents to its gravitational potential~\citep{Diaz:2014kqa,Hartl:2021aio}. In observational studies, however, this method is limited by the availability of only line-of-sight (LoS) velocities for galaxies, as proper motions—especially for distant tracers—are often inaccessible or poorly constrained~\citep{Makarov:2025}. This incomplete view of phase space introduces significant ambiguity in dynamical modeling, as the full 3D motions of galaxies must be inferred from partial data. To deal with this limitation, simplified assumptions about the velocity field are frequently adopted. Two such models are the minor infall and major infall approximations~\citep{Karachentsev:2006ww}. In the minor infall model, it is assumed that the observed LoS velocity of a galaxy accounts for the entirety of its motion relative to the LG's CoM~\citep{Wagner:2025wrp}. This treatment neglects transverse velocity components and tends to underestimate the system's mass by ignoring kinetic energy in unmeasured directions.

By contrast, the major infall model assumes that the motion of the LG’s CoM itself is purely along the line of sight. In this case, the radial velocities of satellites are interpreted under the assumption that the group’s systemic motion affects only the LoS component. This method can lead to overestimates of the LG mass, as it may artificially inflate the inferred velocity dispersion—particularly if unbound tracers or outliers contaminate the sample. Reconciling the biases introduced by these assumptions is crucial for improving the reliability of LG mass estimates. The growing availability of proper motion measurements from instruments like Gaia and HST, as well as next-generation simulations, provides an opportunity to refine our understanding of group dynamics and assess the validity of simplified models.

\begin{figure}
    \centering
    \includegraphics[width=0.9\linewidth]{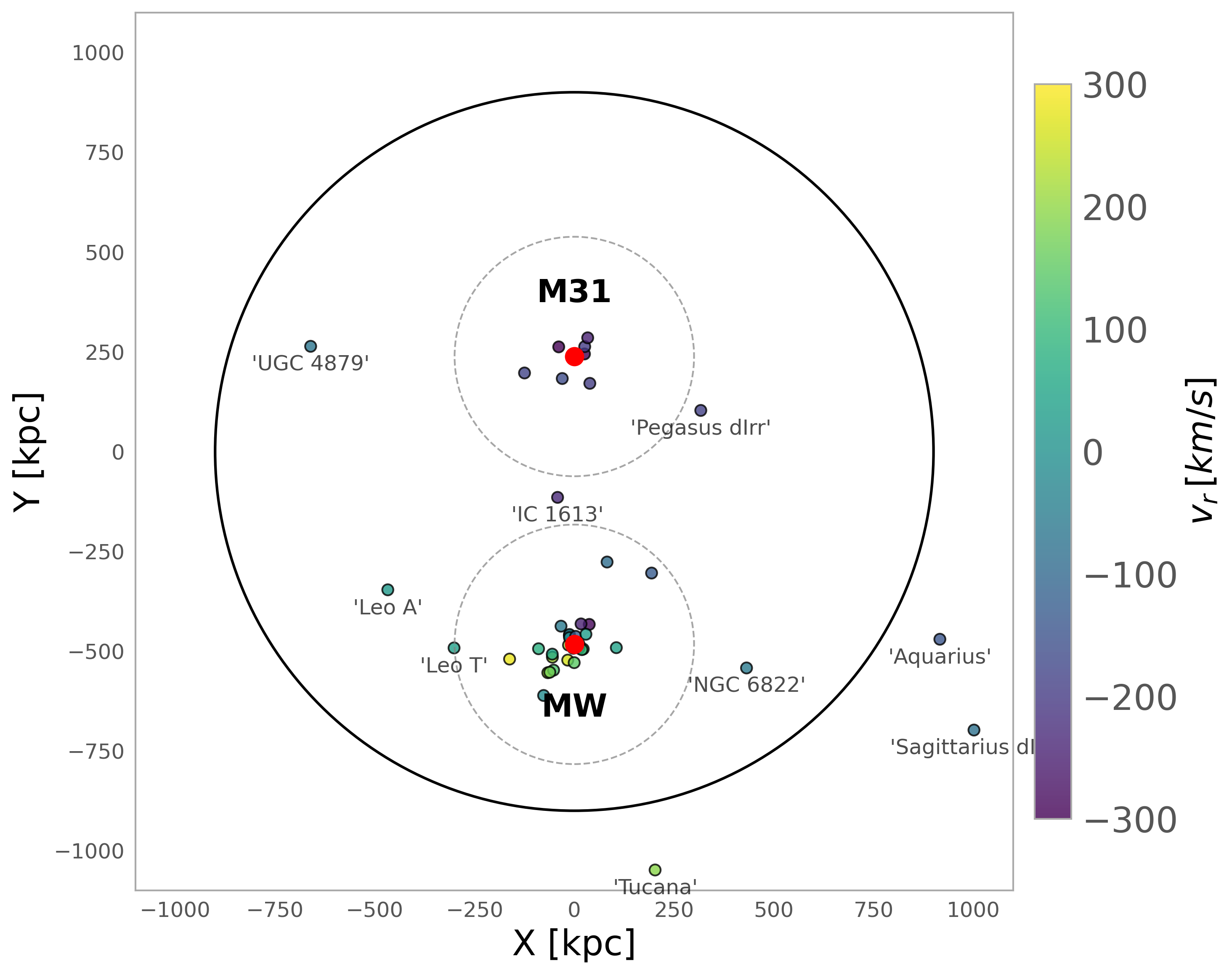}
\caption{  {Distribution of Local Group dwarf galaxies. The sample includes MW and M31 satellites, along with isolated galaxies within 4 Mpc, serving as tracers for LG dynamics. The data is taken from~\citep{2020AJ....160..124M,2020RNAAS...4..229M}.}}
    \label{fig:lg_dwarfs}
\end{figure}

In this work, we employ the IllustrisTNG simulation (see~\cite{bib:Nelson2015}) suite—as a controlled testbed for evaluating virial mass estimators under realistic conditions. We apply these calibrated estimators to the observed kinematics of dwarf galaxies within the velocity surface around the LG CoM. By incorporating geometric corrections and simulation-informed bias mitigation, we obtain a robust mass estimate for the LG that reconciles differences between traditional virial methods, Timing Arguments, and simulation-based approaches. Moreover, we test how the minor and major infall models from the MW like galaxies changes the mass prediction for the Virial Theorem and based on the two predictions we find a robust virial mass estimate.

This paper is structured as follows: Section~\ref{sec:theory} introduces the theoretical foundation for virial mass estimation and details the assumptions underlying different infall models. Section~\ref{sec:simulation} presents the implementation of these models within the TNG50 simulation and the development of bias corrections. In Section~\ref{sec:data}, we apply our methodology to observational data on LG dwarf galaxies. Section~\ref{sec:Dis} discusses the results in the broader context of galaxy group dynamics and cosmological expectations.

\section{Theory}
\label{sec:theory}

\subsection{Relative Motion and Infall Models}

\begin{figure}[t]
\centering
\includegraphics[width=0.85\linewidth]{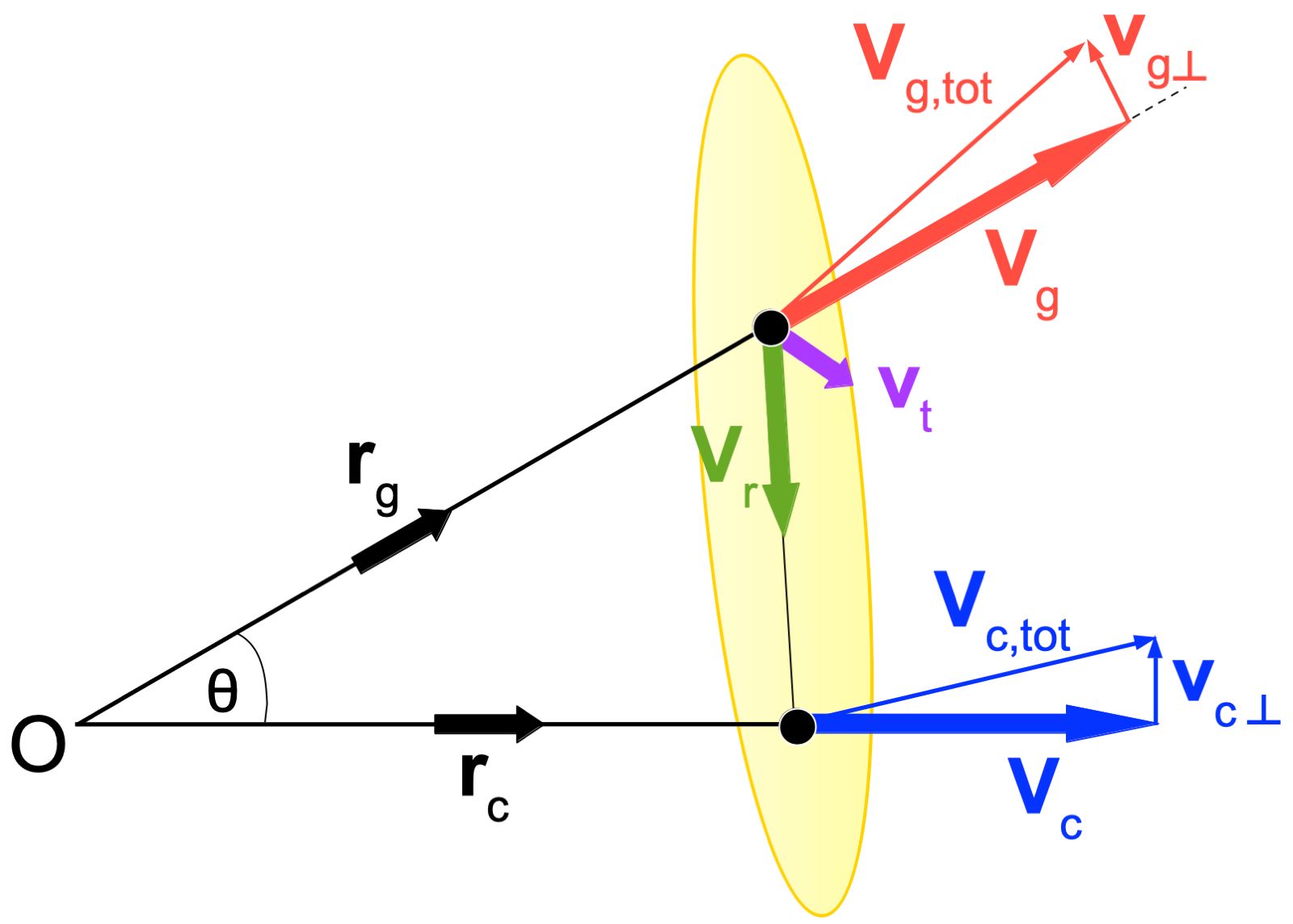}
\caption{  {Relative motion of two galaxies. The position vectors $\mathbf{r}_1$ and $\mathbf{r}_2$ (with distances $r_{\mathrm{g}}$ and $r_{\mathrm{c}}$ from the observer) form an angle $\theta$. Velocity vectors $\mathbf{v}_1$ and $\mathbf{v}_2$ include both radial and tangential components, though only their line-of-sight projections are directly observable.}}
\label{fig:rel_motion}
\end{figure}

To investigate the dynamics of galaxy pairs within the LG, we first transform observed velocities into the (CoM frame. The physical separation $r_{\mathrm{gc}}$ between two galaxies is determined from their angular separation $\theta$ on the sky and their distances $r_{\mathrm{g}}$ and $r_{\mathrm{c}}$ from the observer:
\begin{equation}
r_{\mathrm{gc}}^2 = r_{\mathrm{g}}^2 + r_{\mathrm{c}}^2 - 2 r_{\mathrm{g}} r_{\mathrm{c}} \cos\theta,
\end{equation}
where $r_{\mathrm{g}}$ and $r_{\mathrm{c}}$ denote the distances of the galaxies from the observer (see Fig.~\ref{fig:rel_motion}). Due to observational constraints, only the LoS velocity components $v_{\mathrm{LoS}}$ can be measured directly; the other direction $v_{\perp}$ remain largely unconstrained. As such, inferring the true radial infall velocity relative to the LG’s CoM requires assumptions about the unobserved components. Below, we present two limiting models for estimating these infall velocities, following \cite{2006Ap.....49....3K} and the updated interpretation in \citep{Wagner:2025wrp,Benisty:2025a}.

The minor infall model treats the galaxy and CoM symmetrically and assumes that both have vanishing tangential velocities $v_{\perp,\mathrm{c}} = v_{\perp,g} = 0$. The estimated radial infall velocity $v_{\mathrm{r,min}}$ of galaxy toward the CoM is:
\begin{align}
v_{\mathrm{r,min}} &= \frac{v_{\mathrm{c}} {r}_{\mathrm{c}} + v_{g} {r}_g - \cos\theta_{c,g} \left( v_g {r}_{\mathrm{c}} + v_{\mathrm{c}} {r}_g \right)}{r_{gc}},
\label{eq:v_min}
\end{align}
where $v_{\mathrm{c}}$ and $v_j$ are the LoS velocities of the CoM and the galaxy and $\theta_{c,j}$ is the angular separation between the center and the galaxy. This model assumes negligible non-radial motion, which is plausible for a statistical ensemble but could break down in finely tuned individual systems \citep{Wagner:2025wrp}.
 
In contrast, the major infall model adopts an asymmetric formulation by projecting the velocity difference between the galaxy and CoM onto their separation vector. Assuming vanishing transverse motion $v_{\mathrm{t}} = 0$, the inferred radial infall velocity is:
\begin{align}
v_{\mathrm{r,maj}} &= \frac{v_g - v_{\mathrm{c}} \cos\theta_{c,g}}{{r}_g - {r}_{\mathrm{c}} \cos\theta_{\mathrm{c},g}}r_{gc} ,
\label{eq:v_maj}
\end{align}
This model emphasizes the relative velocity component along the direct line connecting the galaxy to the CoM, implicitly neglecting any perpendicular contributions. Both infall models assume that galaxies act as equal-mass tracers of the LG’s mass distribution. Their contrasting assumptions bracket the range of plausible radial velocities, providing upper and lower bounds that, when combined with virial theorem estimates, help constrain the total mass of the LG.

\subsection{The Virial Theorem}
The virial theorem provides a powerful tool for estimating the gravitational mass of a system using kinematic tracers\citep{Limber:1960,bib:Bahcall1981,Heisler:1985,bib:Evans2011,bib:Tully2015,Benisty:2024tlv}. Consider a system of $N$ test particles (e.g., satellite galaxies or halo stars) with positions $\mathbf{r}_i$ and velocities $\mathbf{v}_i$ orbiting in a gravitational potential $\phi(\mathbf{r})$ generated by an unseen mass distribution $\rho(\mathbf{r})$. Observational limitations often restrict measurements to projected positions and LoS velocities, necessitating simplified mass estimators. The virial theorem:
\begin{equation}
2\langle K \rangle + \langle U_G \rangle = 0, 
\end{equation}
gives:
\begin{equation}
M = \frac{f_{\rm VT}}{G} \frac{\sum_{i} v_{i}^2}{\sum_i R_i^{-1}} = f_{\rm VT} \frac{\sigma_v^2 r_{G}}{G},
\label{eq:VT}
\end{equation}  
where $f_{\rm VT}$ encapsulates assumptions about the tracer geometry and potential, $\sigma_v$ is the one-dimensional velocity dispersion, and $r_G$ represents a characteristic gravitational radius derived from the 3D. 

Application of the virial theorem to observational data requires careful consideration of projection effects. When only LoS velocities and projected distances are available, the three-dimensional estimator adapts by introducing a geometric correction. For a spherically symmetric system with isotropic velocities and sufficient tracers, the virial mass becomes:
\begin{equation}
M = \frac{3\pi}{2} \frac{\sigma_{v,los}^2 R_G}{G},
\end{equation}
where $R_G$ is the projected gravitational radius and $\sigma_{v,los}$ is the dispersion of the LoS velocity. This replaces the three-dimensional $r_G$ with the projected $R_G$ and the radial velocity dispersion.

The LG deviates from this idealized scenario. Dominated by the MW and M31, its binary mass distribution and anisotropic tracer population preclude the use of standard spherical corrections. Furthermore, the limited number of tracers and their asymmetric distribution around two centers (rather than a single halo) violate the assumptions behind the projected virial formula. To address this, cosmological simulations replicating the LG's unique configuration calibrate the factor $f_{\rm VT}$, accounting for its non-trivial geometry and dynamical interactions. Thus, the virial mass estimate combines observed velocities and distances with simulation-derived adjustments to mitigate biases. To assess the role of dark matter in galaxy groups, we model these systems as dominated by one or two giant galaxies (masses $M_1$, $M_2$) with $N$ remote tracer galaxies of average mass $m$. Neglecting self-gravity among the tracers, the gravitational potential energy is dominated by the giant galaxies:
\begin{equation}
 \left\langle U_G  \right\rangle = - G\left( \frac{M_1}{r_{1G}} + \frac{M_2}{r_{2G}} \right).
\end{equation}
Here, $r_{1G}$ and $r_{2G}$ are scale lengths inversely proportional to the tracers' mean distances from $M_1$ and $M_2$, respectively, defined as:
\begin{equation}
r_{G} = N / \sum_{i} \frac{1}{r_{i0}},   
\end{equation}
where $r_{i0}$ denotes the distance of the $i$-th tracer to the giant galaxies' CoM. Applying the virial theorem (Eq.~\ref{eq:VT}) is modified with:
\begin{equation}
\frac{1}{r_{G}} = \frac{\gamma}{r_{1G}} +  \frac{1 - \gamma}{r_{2G}}
\end{equation}
where $\gamma = M_{MW}/M_{\rm Tot}$ is the mass ratio and $M = M_1 + M_2$ is the total mass. By adopting mass ratios $\gamma$ inferred from observations (e.g., $\gamma = 0.7$ for the MW-M31), the total mass can be determined from $\sigma_v$ and the tracers' spatial distribution. 

\begin{figure}[t!]  
\centering  
\includegraphics[width=0.95\linewidth]{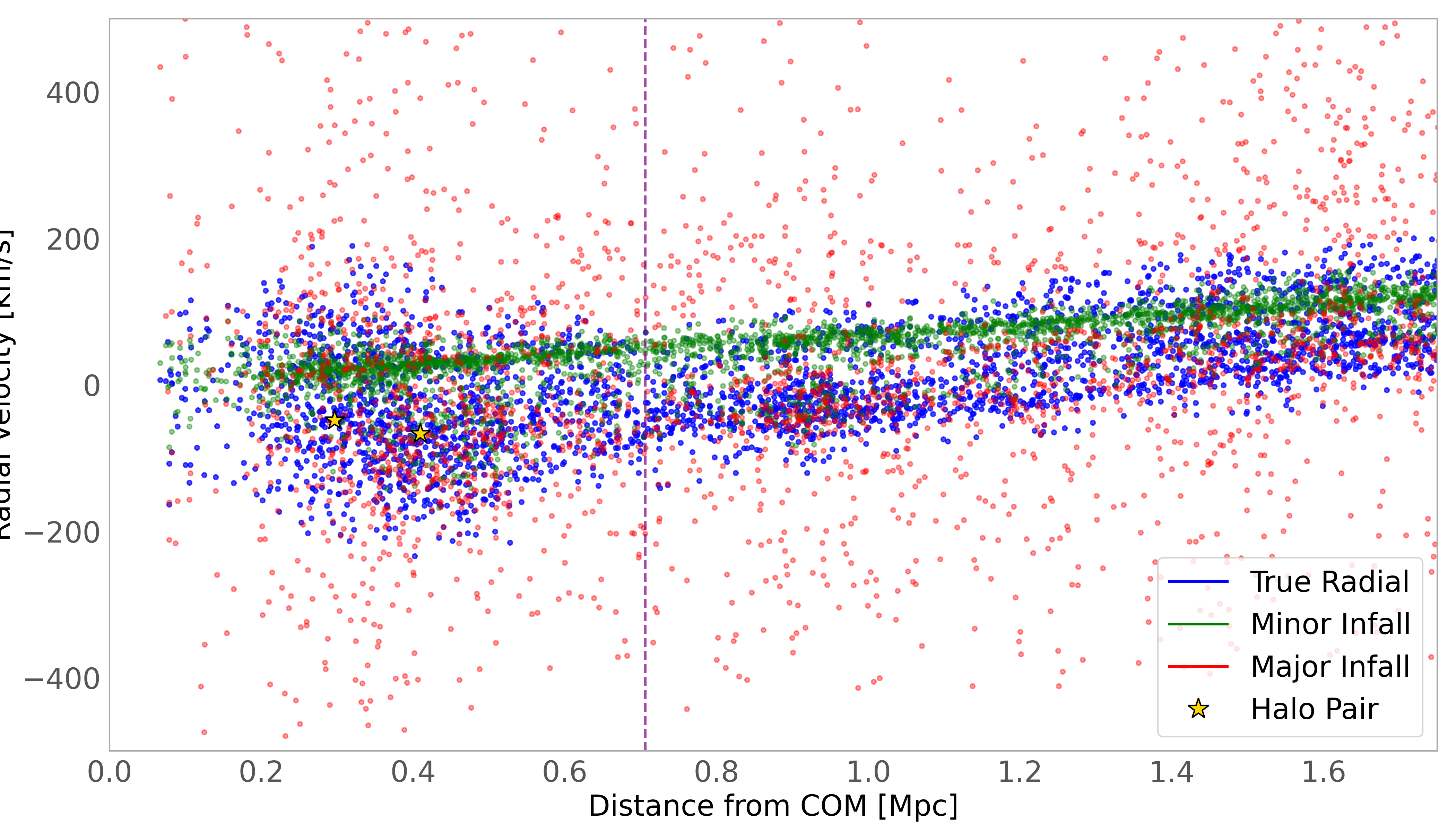}  
\caption{  {Radial velocity distribution for an isolated galaxy pair in TNG-50. Velocities relative to the CoM reveal anisotropic kinematics: minor-infall motions (blue) show suppressed dispersion, while major-infall velocities (red) reflect orbital dynamics. Distances are normalized to the pair separation.}}
\label{fig:tng_LG}  
\vspace{0.5cm}
\includegraphics[width=0.75\linewidth]{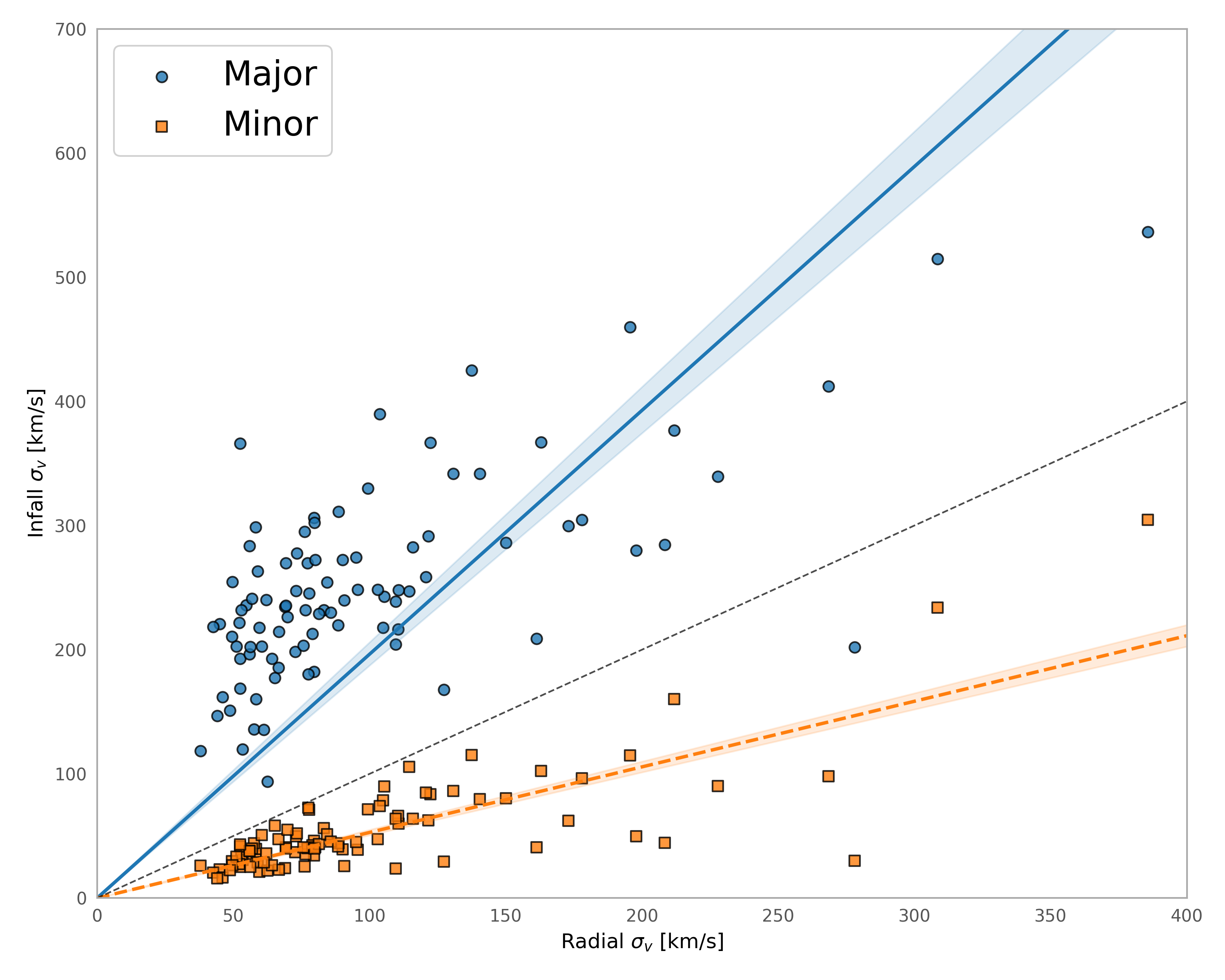} 
\caption{  {Velocity dispersions of bound subhalos around the pair based on the minor and major infall models vs. the true radial velocity. }}  
\label{fig:disp}  
\end{figure}  

\section{N-Body Simulation}
\label{sec:simulation}

The cosmological magnetohydrodynamical simulations analyzed in this work are drawn from the IllustrisTNG project \citep{bib:Vogelsberger2014,bib:Nelson2015}, specifically the TNG-50 run. This simulation models a $\Lambda$CDM universe in a periodic box of side length $51.7~\mathrm{Mpc}$ with a resolution sufficient to resolve galaxy-scale dynamics. The simulation adopts cosmological parameters consistent with Planck 2018 results~\citep{Planck:2018vyg}. To identify LG-like systems, we select isolated galaxy pairs from TNG-50 using the zero-velocity surface criteria~\citep{bib:Sandage1986}. For a halo pair with total mass $M$, the turnaround radius $r_\mathrm{ta}$ is defined as the boundary where inward peculiar velocities balance the Hubble expansion. This radius is calculated via:  
\begin{equation}  
r_\mathrm{ta} = \left( \frac{G M}{t_U^2} \right)^{1/3},  
\label{eq:turnaround_radius}  
\end{equation}  
where $t_U$ is the age of the Universe. Pairs are classified as isolated if no other halos with $M > 10^{10} \, M_\odot$ reside within $r_\mathrm{ta}$. This ensures dynamical isolation akin to the LG, where the MW and M31 dominate the local mass distribution. Candidate pairs are further filtered by mass ratio $\gamma \in [0.5, 1.2]$ and separation $ r \in [0.5, 1.2]\,Mpc$, mirroring the MW-M31 configuration. For each selected pair, we transform ourselves as seen by an observer at the least massive galaxy (the MW-type) into the CoM of the pair, considering the minor and major infall from the MW-like halo.

Fig.~\ref{fig:tng_LG} compares the radial velocity distributions predicted by the minor and major infall models to the true velocities for a representative pair. While neither model reproduces the exact radial velocity toward the CoM, they yield distinct distributions: the minor infall model exhibits a narrower spread, while the major infall model predicts a broader distribution relative to the true velocities. To quantify these differences Fig.~\ref{fig:disp} shows the velocity dispersion profiles of bound subhalos for both models. The minor infall model systematically underestimates the velocity dispersion (biased low), whereas the major infall model overestimates it (biased high). These opposing biases suggest that the true mass lies between the estimates derived from the two models.

The velocity anisotropy—deviations from isotropic orbital motions—necessitates a geometric correction factor, $ f_{\mathrm{VT}}$, to reconcile VT predictions with observations. Applying the VT to dwarf galaxies with known enclosed masses within the turnaround radius, we calculate $f_{\mathrm{VT}}$ for each infall scenario. The major infall model’s overestimated dispersion $\sigma_v$ requires a correction of $f_{\mathrm{VT,Maj}} = 0.7$ to match the true mass $M_\mathrm{True}$. Conversely, the minor infall model’s underestimated dispersion necessitates a significantly larger correction, $f_{\mathrm{VT,Min}} = 16.8$. This stark contrast highlights the minor model’s poor agreement with observations without correction, whereas the major model aligns more closely with expectations. Both corrections are consistent with analytic models for isolated halos, underscoring the sensitivity of virial mass estimates to velocity anisotropy.

\begin{figure}[t]  
\centering  
\includegraphics[width=0.9\linewidth]{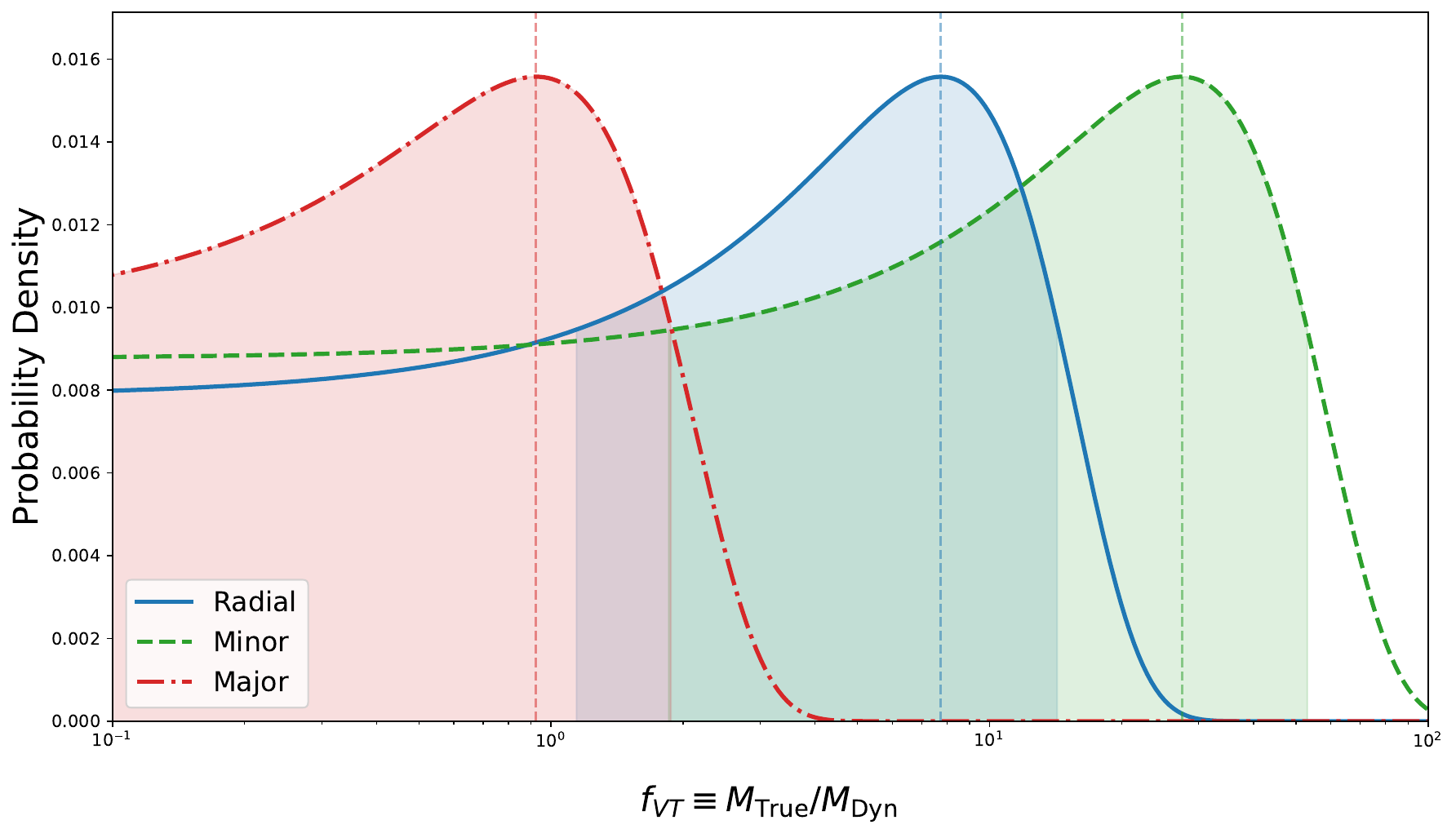}  
\caption{Probability density distributions of the virial theorem correction factor $f_{\rm VT} \equiv M_{\rm True}/M_{\rm Dyn}$, where $M_{\rm Dyn} \equiv \sigma_v^2 r_G / G$, calibrated using the TNG50 simulation. The curves correspond to three dynamical models: Radial (uncorrected velocity dispersion, Minor Infall and Major Infall. The peaks reflect the optimal correction factors derived from simulation ensembles, with widths indicating uncertainties in velocity anisotropy modeling.}  
\label{fig:f_vt}  
\end{figure}

\section{LG Dwarf Galaxies Data}
\label{sec:data}

\begin{figure}[t]  

\includegraphics[width=\linewidth]{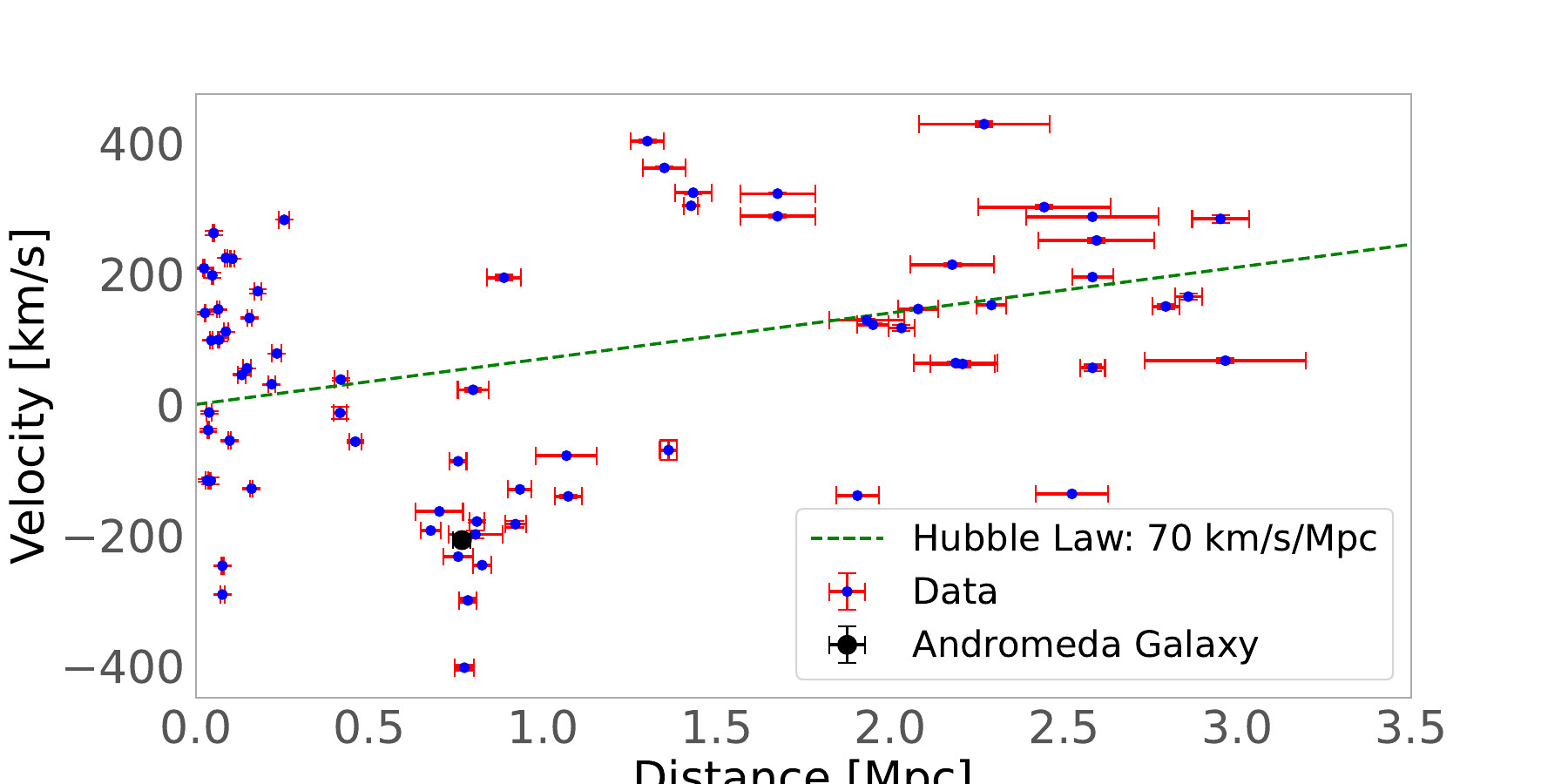}  
\includegraphics[width=\linewidth]{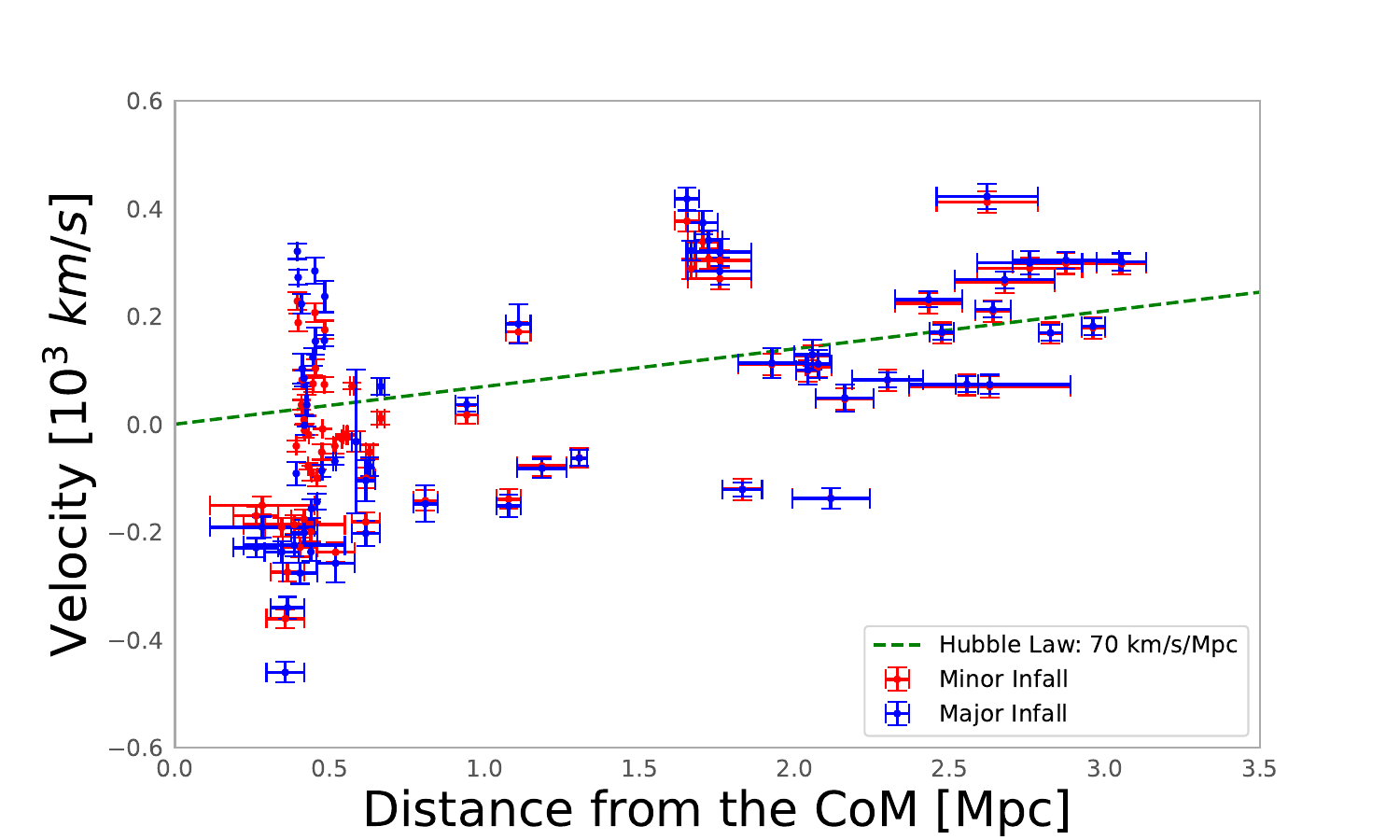}  
\caption{\underline{Upper:} LoS velocity vs. distance from the MW. Nearby galaxies (black points) exhibit deviations from the Hubble Law (green dashed line). \underline{Lower:} Radial velocity vs. distance from the CoM. Velocities are transformed to the CoM frame using $\gamma = 0.7$. The bounded region ($<1\,\text{Mpc}$) contains MW (blue) and M31 (red) satellites.} 
\label{fig:vel_dis_com}  
\end{figure} 

We analyze dwarf galaxies within a 3 Mpc radius around the LG incorporating a comprehensive dataset compiled from \cite{2012AJ....144....4M,2020AJ....160..124M,2020RNAAS...4..229M,Tully:2022rbj}\footnote{\url{https://www.cadc-ccda.hia-iha.nrc-cnrc.gc.ca/en/community/nearby/}}. To determine the LG's CoM, we adopt the observed separation and radial velocity between the MW and M31 from \cite{vanderMarel:2012xp}:

\begin{equation}
r_0 = 0.77 \pm 0.04\,\text{Mpc}, \quad v_{\text{rad}} = -109.3 \pm 4.4\,\text{km/s}.
\end{equation}

The CoM is assumed to lie along the line connecting the MW and M31, with its precise location set by the mass ratio $ M_{\mathrm{MW}} / M_{\mathrm{LG}}$. Following \cite{Diaz:2014kqa}, we adopt $\gamma = 0.7$, consistent with recent dynamical studies of the LG. Fig.~(\ref{fig:vel_dis_com}) illustrates the radial velocity–distance relation for galaxies in our sample. The upper panel shows line-of-sight velocities relative to the MW, which deviate significantly from the linear Hubble flow, reflecting the gravitational influence of the LG. After transforming velocities to the LG CoM frame (lower panel) using the minor and major infall models, the distribution aligns more clearly with expectations from infall models, particularly within the turnaround radius of approximately 1 Mpc. Within this region, two dominant populations emerge: bound satellites of the MW (blue) and M31 (red). Beyond the turnaround radius, the galaxies transition to the Hubble flow regime, highlighting the balance between gravitational collapse and cosmological expansion. 

\begin{figure}[t!]
    \centering
\includegraphics[width=1\linewidth]{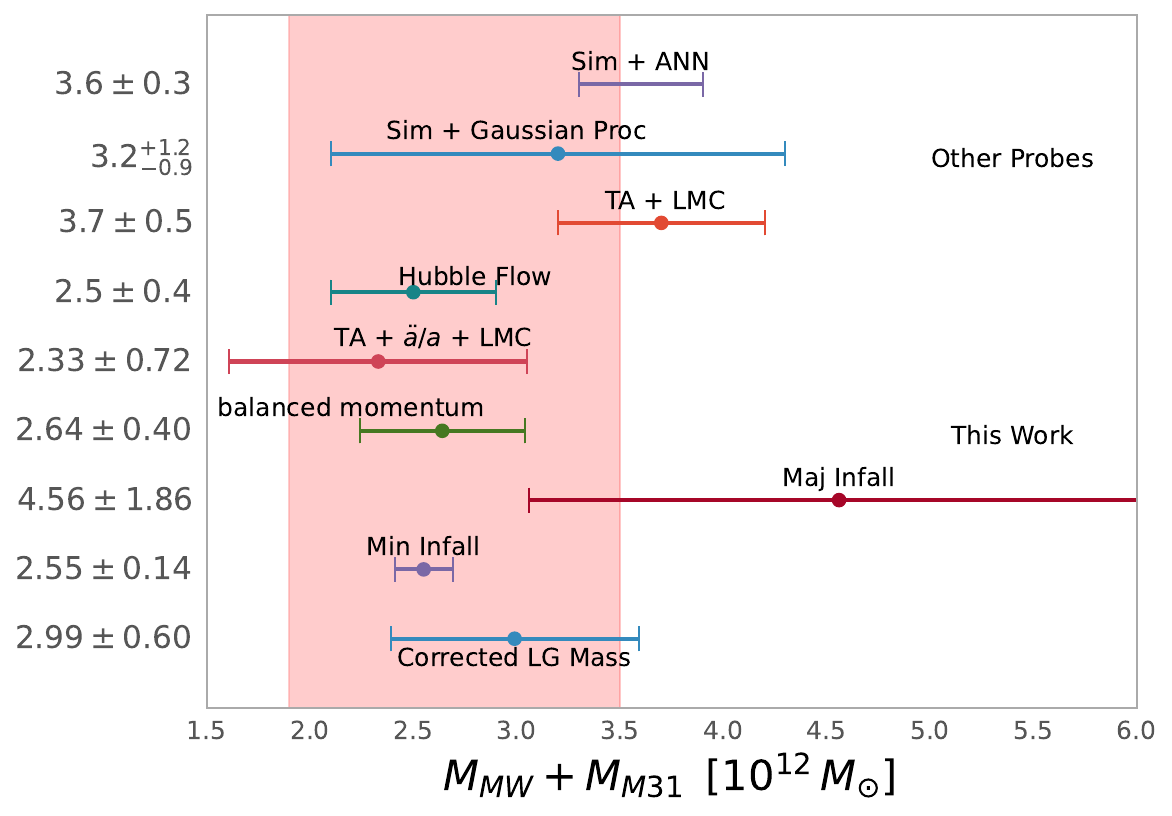}
\caption{Estimates of the total mass of the LG from various methods. The lower entries represent new estimates based on anisotropic infall analysis: a corrected LG mass incorporating projection effects ($2.99 \pm 0.60 \times 10^{12}\,M_\odot$), a minor infall direction ($2.55 \pm 0.14 \times 10^{12}\,M_\odot$), and a major infall direction ($4.56 \pm 1.86 \times 10^{12}\,M_\odot$). These are followed by the balanced momentum method and a revised timing argument with $\ddot{a}/a$ and LMC correction. Top entries include alternative probes such as the Hubble flow, the Timing Argument with Gaussian processes, and artificial neural networks. The shaded red band highlights the range $[1.9$–$3.5] \times 10^{12}\,M_\odot$, consistent with multiple independent analyses. The other probes were taken from~\citep{Diaz:2014kqa,Penarrubia:2014oda,Benisty:2022ive,Benisty:2024lsz,McLeod:2016bjk,Sawala:2022ayk,Sawala:2023sec}.}
    \label{fig:post_mass}
\end{figure} 

To estimate the LG's dynamical mass, we examine galaxy velocity dispersions along the principal infall axes. For the minor-infall, characterized by suppressed vertical motions, we measure a velocity dispersion of $133.44 \pm 29.76\,\text{km/s}$. For the major-axis infall, dominated by orbital streaming, the dispersion increases to $178.67 \pm 36.21\,\text{km/s}$. Adopting a gravitational radius of $r_G = 210 \pm 10\,\text{kpc}$, we apply the virial theorem to infer the enclosed dynamical masses. Using the minor-infall dispersion, the virial mass is:
\begin{equation}
M_{\text{Min}} = \left(2.55 \pm 0.14\right) \times 10^{12}\, M_\odot,
\end{equation}
while the major-infall analysis yields:
\begin{equation}
M_{\text{Maj}} = \left(4.56 \pm 1.86\right) \times 10^{12}\, M_\odot.
\end{equation}
These estimates span a range consistent with anisotropic infall models, where projection effects and tracer contamination bias the inferred mass depending on the viewing infall. Using the correction factors probed by the simulation the LG mass is marginalized to be:
\begin{equation}
M_{\mathrm{LG}} = (2.99 \pm 0.60) \times 10^{12}\, M_\odot,
\end{equation}
which also agrees with the theoretical expectation from the maximum turnaround radius, derived via the velocity surface estimate of~\citep{Karachentsev:2008st}.

The virial mass estimate of $M_{\rm LG} = (2.5 \pm 0.4) \times 10^{12}\,M_\odot$ derived by \cite{Diaz:2014kqa} aligns closely with our minor-infall model prediction ($M_{\rm Min} = 2.55 \times 10^{12}\,M_\odot$). While \cite{Diaz:2014kqa} does not explicitly frame their analysis in terms of "minor infall," their methodology—projecting LoS velocities onto the axis connecting individual galaxies to the LG’s CoM, a hallmark of the minor-infall approximation. This approach systematically underestimates the total mass by excluding unobserved velocity components perpendicular to the CoM-galaxy line. In contrast, our major-infall model, which interprets LoS velocities under the assumption of purely radial motion relative to the CoM, yields a significantly larger mass ($M_{\rm Maj} = 4.56 \times 10^{12}\,M_\odot$). By calibrating these extremes against TNG50 simulations, we reconcile the two regimes into a unified mass estimate ($M_{\rm LG} = 2.99 \times 10^{12}\,M_\odot$), resolving prior inconsistencies and underscoring the critical role of velocity projection corrections in dynamical studies.

\section{Discussion and Conclusions}
\label{sec:Dis}

By analyzing the LG mass budget through cosmological simulations, we test the validity of VT estimators based on minor and major infall models. These models reflect distinct geometric anisotropies in the velocity distribution of dwarf galaxies: the minor infall model, which suppresses vertical motions akin to minor-axis kinematics, systematically underestimates the total mass $M \sim 2.5 \times 10^{12} \, M_\odot$ due to incomplete sampling of the velocity dispersion. Conversely, the major infall model, which incorporates broader velocity distributions aligned with the major axis, overestimates the mass $M \sim 5 \times 10^{12} \, M_\odot$ by including unbound tracers. To reconcile these biases, we derive geometric correction factors that scale the velocity dispersions to match the true enclosed mass. Our fiducial estimate of $M_\mathrm{LG} = 3 \times 10^{12} \, M_\odot$, intermediate between the two extremes, aligns with independent methods and resolves the tension inherent in traditional VT-based approaches.

The fiducial mass also agrees with the TA for the Milky Way–M31 orbit, yielding $M \sim 2.5\text{--}3.2 \times 10^{12} \, M_\odot $ \citep{vanderMarel:2012xp,Garrison-Kimmel:2014,Partridge:2013dsa,Benisty:2024lsz}, as well as with the Hubble Flow fit~\cite{Karachentsev:2008st,Penarrubia:2014oda} and the projected mass method, which gives $\left( 2.34 \pm 0.41\right) \times 10^{12}\,M_{\odot} $~\citep{Makarov:2025}. Although the latter yields a slightly lower biased mass estimate $\sim 2 \times 10^{12}\,M_{\odot}$, it remains consistent with our result within uncertainties. The broader landscape of LG mass estimates also converges around $3 \times 10^{12} \, M_\odot$, with MW and M31 rotation curve studies \citep{Wang:2019ubx, Bobylev:2023} indicating total masses in the range of $[2.8\text{--}3.3] \times 10^{12} \, M_\odot$. Independent mass estimates for the MW ($ [0.9\text{--}1.3] \times 10^{12}\,M_\odot$ ~\cite{Wang:2019ubx}) and M31 ($[1.0\text{--}2.0] \times 10^{12}\,M_\odot$ ~\cite{Sawala:2022ayk}) also support a total mass range of $ [1.9\text{--}3.5] \times 10^{12}\,M_\odot$, as illustrated in Fig.~\ref{fig:post_mass}. The remaining scatter likely results from systematics, including incomplete tracer samples, non-equilibrium dynamics, and the LG’s asymmetric mass distribution, which affects the virial mass estimation factor.

Future progress will require precise proper motions from missions like Gaia and JWST, improved modeling of the Local Volume’s tidal field, and simulations explicitly incorporating nearby galaxy groups. Our findings underscore the necessity of geometric corrections in virial theorem applications and provide a framework for studying anisotropic systems beyond the LG. The LG’s total mass is robustly constrained to $3 \times 10^{12} \, M_\odot$, a value that harmonizes methodological discrepancies, aligns with $\Lambda$CDM predictions, and cements the LG’s role as a cornerstone for testing galaxy formation and dark matter models on group scales.  

\begin{acknowledgements}
DB is supported by a Minerva Fellowship of the Minerva Stiftung Gesellschaft für die Forschung mbH. DFM thanks the Research Council of Norway for their support and the resources provided by UNINETT Sigma2 -- the National Infrastructure for High-Performance Computing and Data Storage in Norway. 
\end{acknowledgements}

\bibliographystyle{aa} 

\bibliography{refences.bib}

\begin{thebibliography}{44}
\expandafter\ifx\csname natexlab\endcsname\relax\def\natexlab#1{#1}\fi

\bibitem[{Aghanim {et~al.}(2020)}]{Planck:2018vyg}
Aghanim, N. {et~al.} 2020, Astron. Astrophys., 641, A6, [Erratum: Astron.Astrophys. 652, C4 (2021)]

\bibitem[{{An} \& {Evans}(2011)}]{bib:Evans2011}
{An}, J.~H. \& {Evans}, N.~W. 2011, \mnras, 413, 1744

\bibitem[{{Bahcall} \& {Tremaine}(1981)}]{bib:Bahcall1981}
{Bahcall}, J.~N. \& {Tremaine}, S. 1981, \apj, 244, 805

\bibitem[{Benisty(2024)}]{Benisty:2024lsz}
Benisty, D. 2024, Astron. Astrophys., 689, L1

\bibitem[{Benisty \& Capozziello(2023)}]{Benisty:2023ofi}
Benisty, D. \& Capozziello, S. 2023, Phys. Dark Univ., 39, 101175

\bibitem[{Benisty {et~al.}(2024)Benisty, Chaichian, \& Tureanu}]{Benisty:2024tlv}
Benisty, D., Chaichian, M.~M., \& Tureanu, A. 2024, Phys. Lett. B, 858, 139033

\bibitem[{Benisty \& Davis(2022)}]{Benisty:2021cmq}
Benisty, D. \& Davis, A.-C. 2022, Phys. Rev. D, 105, 024052

\bibitem[{{Benisty} {et~al.}(2025){Benisty}, {Libeskind}, \& {Hoffman}}]{Benisty:2025a}
{Benisty}, D., {Libeskind}, N.~I., \& {Hoffman}, Y. 2025, {To be Published}

\bibitem[{{Benisty} {et~al.}(2022){Benisty}, {Vasiliev}, {Evans}, {Davis}, {Hartl}, \& {Strigari}}]{Benisty:2022ive}
{Benisty}, D., {Vasiliev}, E., {Evans}, N.~W., {et~al.} 2022, \apjl, 928, L5

\bibitem[{{Bobylev} \& {Bajkova}(2023)}]{Bobylev:2023}
{Bobylev}, V.~V. \& {Bajkova}, A.~T. 2023, Publications of the Pulkovo Observatory, 228, 57

\bibitem[{Chamberlain {et~al.}(2023)Chamberlain, Price-Whelan, Besla, Cunningham, Garavito-Camargo, Pe\~narrubia, \& Petersen}]{Chamberlain:2022fqr}
Chamberlain, K., Price-Whelan, A.~M., Besla, G., {et~al.} 2023, Astrophys. J., 942, 18

\bibitem[{Diaz {et~al.}(2014)Diaz, Koposov, Irwin, Belokurov, \& Evans}]{Diaz:2014kqa}
Diaz, J.~D., Koposov, S.~E., Irwin, M., Belokurov, V., \& Evans, N.~W. 2014, Mon. Not. Roy. Astron. Soc., 443, 1688

\bibitem[{{Garrison-Kimmel} {et~al.}(2014){Garrison-Kimmel}, {Boylan-Kolchin}, {Bullock}, \& {Kirby}}]{Garrison-Kimmel:2014}
{Garrison-Kimmel}, S., {Boylan-Kolchin}, M., {Bullock}, J.~S., \& {Kirby}, E.~N. 2014, \mnras, 444, 222

\bibitem[{Hartl \& Strigari(2022)}]{Hartl:2021aio}
Hartl, O.~V. \& Strigari, L.~E. 2022, Mon. Not. Roy. Astron. Soc., 511, 6193

\bibitem[{Hartl \& Strigari(2025)}]{Hartl:2024bfc}
Hartl, O.~V. \& Strigari, L.~E. 2025, Mon. Not. Roy. Astron. Soc., 539, 160

\bibitem[{{Heisler} {et~al.}(1985){Heisler}, {Tremaine}, \& {Bahcall}}]{Heisler:1985}
{Heisler}, J., {Tremaine}, S., \& {Bahcall}, J.~N. 1985, \apj, 298, 8

\bibitem[{{Karachentsev} \& {Kashibadze}(2006)}]{2006Ap.....49....3K}
{Karachentsev}, I.~D. \& {Kashibadze}, O.~G. 2006, Astrophysics, 49, 3

\bibitem[{Karachentsev {et~al.}(2009)Karachentsev, Kashibadze, Makarov, \& Tully}]{Karachentsev:2008st}
Karachentsev, I.~D., Kashibadze, O.~G., Makarov, D.~I., \& Tully, R.~B. 2009, Mon. Not. Roy. Astron. Soc., 393, 1265

\bibitem[{Karachentsev {et~al.}(2007)}]{Karachentsev:2006ww}
Karachentsev, I.~D. {et~al.} 2007, Astron. J., 133, 504

\bibitem[{Lemos {et~al.}(2021)Lemos, Jeffrey, Whiteway, Lahav, Libeskind, \& Hoffman}]{Lemos:2020vhj}
Lemos, P., Jeffrey, N., Whiteway, L., {et~al.} 2021, Phys. Rev. D, 103, 023009

\bibitem[{Li \& White(2008)}]{Li:2007eg}
Li, Y.-S. \& White, S. D.~M. 2008, Mon. Not. Roy. Astron. Soc., 384, 1459

\bibitem[{{Limber} \& {Mathews}(1960)}]{Limber:1960}
{Limber}, D.~N. \& {Mathews}, W.~G. 1960, \apj, 132, 286

\bibitem[{{Makarov} {et~al.}(2025){Makarov}, {Makarov}, {Kozyrev}, \& {Libeskind}}]{Makarov:2025}
{Makarov}, D., {Makarov}, D., {Kozyrev}, K., \& {Libeskind}, N. 2025, arXiv e-prints, arXiv:2503.12612

\bibitem[{{McConnachie}(2012)}]{2012AJ....144....4M}
{McConnachie}, A.~W. 2012, \aj, 144, 4

\bibitem[{{McConnachie} \& {Venn}(2020{\natexlab{a}})}]{2020AJ....160..124M}
{McConnachie}, A.~W. \& {Venn}, K.~A. 2020{\natexlab{a}}, \aj, 160, 124

\bibitem[{{McConnachie} \& {Venn}(2020{\natexlab{b}})}]{2020RNAAS...4..229M}
{McConnachie}, A.~W. \& {Venn}, K.~A. 2020{\natexlab{b}}, Research Notes of the American Astronomical Society, 4, 229

\bibitem[{McLeod \& Lahav(2020)}]{McLeod:2019cfg}
McLeod, M. \& Lahav, O. 2020, JCAP, 09, 056

\bibitem[{McLeod {et~al.}(2017)McLeod, Libeskind, Lahav, \& Hoffman}]{McLeod:2016bjk}
McLeod, M., Libeskind, N., Lahav, O., \& Hoffman, Y. 2017, JCAP, 12, 034

\bibitem[{{Nelson} {et~al.}(2015){Nelson}, {Pillepich}, {Genel}, {Vogelsberger}, {Springel}, {Torrey}, {Rodriguez-Gomez}, {Sijacki}, {Snyder}, {Griffen}, {Marinacci}, {Blecha}, {Sales}, {Xu}, \& {Hernquist}}]{bib:Nelson2015}
{Nelson}, D., {Pillepich}, A., {Genel}, S., {et~al.} 2015, Astronomy and Computing, 13, 12

\bibitem[{Partridge {et~al.}(2013)Partridge, Lahav, \& Hoffman}]{Partridge:2013dsa}
Partridge, C., Lahav, O., \& Hoffman, Y. 2013, Mon. Not. Roy. Astron. Soc., 436, 45

\bibitem[{Pe\~narrubia {et~al.}(2016)Pe\~narrubia, G\'omez, Besla, Erkal, \& Ma}]{Penarrubia:2015hqa}
Pe\~narrubia, J., G\'omez, F.~A., Besla, G., Erkal, D., \& Ma, Y.-Z. 2016, Mon. Not. Roy. Astron. Soc., 456, L54

\bibitem[{Pe\~narrubia {et~al.}(2014)Pe\~narrubia, Ma, Walker, \& McConnachie}]{Penarrubia:2014oda}
Pe\~narrubia, J., Ma, Y.-Z., Walker, M.~G., \& McConnachie, A. 2014, Mon. Not. Roy. Astron. Soc., 443, 2204

\bibitem[{{Sandage}(1986)}]{bib:Sandage1986}
{Sandage}, A. 1986, \apj, 307, 1

\bibitem[{Sawala {et~al.}(2023{\natexlab{a}})Sawala, Pe\~narrubia, Liao, \& Johansson}]{Sawala:2023sec}
Sawala, T., Pe\~narrubia, J., Liao, S., \& Johansson, P.~H. 2023{\natexlab{a}}, Mon. Not. Roy. Astron. Soc., 526, L77

\bibitem[{Sawala {et~al.}(2023{\natexlab{b}})Sawala, Teeriaho, \& Johansson}]{Sawala:2022ayk}
Sawala, T., Teeriaho, M., \& Johansson, P.~H. 2023{\natexlab{b}}, Mon. Not. Roy. Astron. Soc., 521, 4863

\bibitem[{{Tully}(2015)}]{bib:Tully2015}
{Tully}, R.~B. 2015, \aj, 149, 54

\bibitem[{Tully {et~al.}(2023)}]{Tully:2022rbj}
Tully, R.~B. {et~al.} 2023, Astrophys. J., 944, 94

\bibitem[{{van den Bergh}(1999)}]{Bergh1999}
{van den Bergh}, S. 1999, \aapr, 9, 273

\bibitem[{van~der Marel {et~al.}(2012)van~der Marel, Fardal, Besla, Beaton, Sohn, Anderson, Brown, \& Guhathakurta}]{vanderMarel:2012xp}
van~der Marel, R.~P., Fardal, M., Besla, G., {et~al.} 2012, Astrophys. J., 753, 8

\bibitem[{van~der Marel \& Guhathakurta(2008)}]{vanderMarel:2007yw}
van~der Marel, R.~P. \& Guhathakurta, P. 2008, Astrophys. J., 678, 187

\bibitem[{{Vogelsberger} {et~al.}(2014){Vogelsberger}, {Genel}, {Springel}, {Torrey}, {Sijacki}, {Xu}, {Snyder}, {Bird}, {Nelson}, \& {Hernquist}}]{bib:Vogelsberger2014}
{Vogelsberger}, M., {Genel}, S., {Springel}, V., {et~al.} 2014, \nat, 509, 177

\bibitem[{Wagner \& Benisty(2025)}]{Wagner:2025wrp}
Wagner, J. \& Benisty, D. 2025 [\eprint[arXiv]{2501.13149}]

\bibitem[{Wang {et~al.}(2020)Wang, Han, Cautun, Li, \& Ishigaki}]{Wang:2019ubx}
Wang, W., Han, J., Cautun, M., Li, Z., \& Ishigaki, M.~N. 2020, Sci. China Phys. Mech. Astron., 63, 109801

\bibitem[{Wempe {et~al.}(2024)Wempe, Lavaux, White, Helmi, Jasche, \& Stopyra}]{Wempe:2024rfj}
Wempe, E., Lavaux, G., White, S. D.~M., {et~al.} 2024, Astron. Astrophys., 691, A348

\end{thebibliography}

\end{document}